\documentclass[conf]{IEEEtran}

\usepackage[left=0.75in,right=0.75in,top=0.75in,bottom=0.75in]{geometry}

\usepackage{color}
\usepackage{qtree}
\usepackage{times}
\usepackage{epsfig}
\usepackage{graphicx}
\usepackage{epstopdf}
\usepackage{algpseudocode}
\usepackage{algorithm}
\usepackage{amsmath}
\usepackage{amssymb}
\usepackage{amsxtra}
\usepackage{multirow}
\usepackage{mathtools}
\usepackage{cleveref}
\usepackage{url}
\usepackage{cite}
\usepackage{amsthm}
\usepackage{adjustbox}
\usepackage[latin1]{inputenc}
\usepackage{caption}
\usepackage[caption = false]{subfig}
\usepackage{float}

\newtheorem*{proof*}{Proof}

\begin{document}
%
\title{Modeling and Assessment of IoT Supply Chain Security Risks: The Role of Structural and Parametric Uncertainties}
%
%
%
%

\author{ \IEEEauthorblockN{\large Timothy Kieras}, \IEEEauthorblockN{\large Muhammad Junaid Farooq}, and \IEEEauthorblockN{\large Quanyan Zhu} \\ \IEEEauthorblockA{Department of Electrical \& Computer Engineering, Tandon School of Engineering, \\New York University, Brooklyn, NY 11201, USA,} Emails: \{tk2375, mjf514, qz494\}@nyu.edu. \vspace{-0.0in}
\thanks{\vspace{-0.0in}\hrule \vspace{0.2cm}
This research is partially supported by award 2015-ST-061-CIRC01, U. S. Department of Homeland Security, awards ECCS-1847056 and SES-1541164 from National Science of Foundation (NSF).}
}
\maketitle

\begin{abstract}
Supply chain security threats pose new challenges to security risk modeling techniques for complex ICT systems such as the IoT. With established techniques drawn from attack trees and reliability analysis providing needed points of reference, graph-based analysis can provide a framework for considering the role of suppliers in such systems. We present such a framework here while highlighting the need for a component-centered model. Given resource limitations when applying this model to existing systems, we study various classes of uncertainties in model development, including structural uncertainties and uncertainties in the magnitude of estimated event probabilities. Using case studies, we find that structural uncertainties constitute a greater challenge to model utility and as such should receive particular attention. Best practices in the face of these uncertainties are proposed.
\end{abstract}


%
\IEEEpeerreviewmaketitle

\section{Introduction}

Information and communications technology (ICT) systems are becoming increasingly complex, consisting of various different components connected together~\cite{implications}. Often, these components are manufactured, controlled, or operated by different entities in different regions of the world. It is almost impossible to have centralized control over all entities in the supply chain. Therefore, in addition to the conventional risks of system failures, there is another layer of risks emanating from supply chain actors. With the proliferation of Internet of Things (IoT) devices and networks, these risks are further amplified due to an unregulated and extremely heterogeneous ecosystem. Hence, it is becoming critical to develop methodologies to measure and analyze these supply chain risks. A more alarming alarming concern is that the IoT systems directly interact with critical infrastructure systems leading to the possibility of cascaded failures and other disastrous consequences.

Supply chain risk analysis in IoT systems is a challenge due to the lack of direct applicability of traditional methodologies in supply chain risk management~\cite{wf_iot}. This challenge is primarily due to the structural complexity of the systems and the various different uncertainties that impact the assessment of systemic risk. Since fundamental design methodologies in technical industries rely heavily on modularity and abstraction, black-box systems that are difficult, if not impossible, to audit 
pose a constraint within which critical infrastructure security must be pursued. Processes to verify that components have been manufactured without hidden security vulnerabilities may be strategically important; however, the evident specialization required to produce complex ICT systems entails the need for a comparable specialization to provide credible verification. In general, then, acquirers or integrators of complex components must be prepared to assess and manage risks associated with supplier trust~\cite{scrm}. This challenging task calls for ongoing development of risk assessment methodologies and heuristics for evaluating the trustworthiness of suppliers~\cite{NIST, mitre}.

\begin{figure}[t]
    \centering
    \includegraphics[width=.35\textwidth]{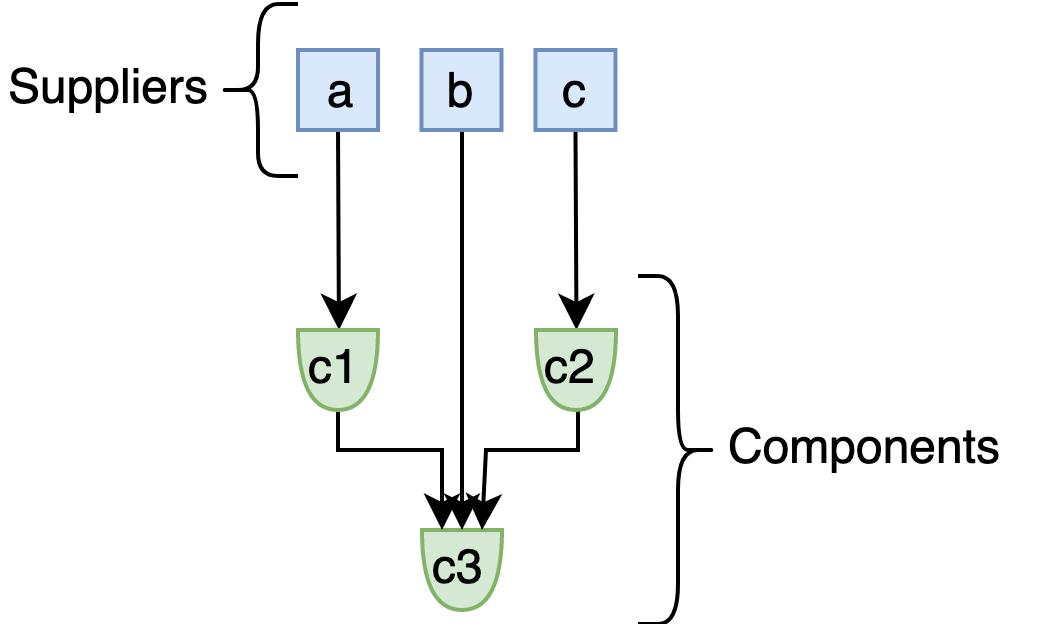}
    \caption{A simple system graph depicting three components, their suppliers, and dependencies between them.}
    \label{fig:simple_system}
\end{figure}

When developing modeling practices suited to assessing these risks, it is necessary to consider the operational constraints within which model development proceeds. Given the uncertainties inevitably present in model development, processes should be designed to prioritize the clarification of uncertainties that pose the greatest threat to model utility. In this paper we present a brief graph-based model suited for the assessment of supply chain risks, and develop case studies that illustrate the kinds of uncertainties that may impede the application of the modeling process to a given system. We then propose that structural uncertainties have the potential to be more significant than uncertainties in the probabilities of basic events. Similarly, among structural uncertainties, those pertaining to higher level components are particularly critical. Accordingly, the principle is developed that supply chain security risk assessment should prioritize the accurate structural assessment of systems in order to improve the reliability of risk modeling in practice.

The rest of the paper is organised as follows. Section II provides an overview of the system model, including an explanation of the risk analysis enabled by the model. Section III outlines two kinds of uncertainties that may affect the utility of the model in practice. Section IV develops a series of case studies based on these uncertainties, illustrating important aspects of model performance when subject to uncertainty and pointing to priorities when seeking to mitigate the effect of these uncertainties. In conclusion we offer recommendations based on these case studies.

\section{System Model}


The model developed here can be introduced by a comparison with attack trees ~\cite{mauw2005foundations,jha2002two}. An attack tree is a directed acyclic graph (DAG) based security analysis technique relying on the fact that security incident analysis is very similar to classical reliability analysis ~\cite{kordy2014dag, xiong2019threat, RN06973090519990101, amoroso_fundamentals_1994}. In an attack tree, nodes represent events or sub-goals that contribute to the accomplishing of a central event or goal that is the desired aim of the attack under analysis. Edges connect nodes so that resulting paths from leaf nodes to the top event represent feasible, completed attacks. Given probabilities of these events, it is possible to compute the likelihood of each path and so consider optimal detection and mitigation choices. While recent research has explored the automated construction of attack graphs on systems by interfacing network scanning applications and databases of known vulnerabilities, it remains the case that the most important elements in attack graphs are essentially discovered through empirical investigation and expert knowledge of a system ~\cite{sheyner2002automated, ou2006scalable}. It is the discovery of the causal connection between one event and another that forms the basis of the utility of an attack graph. This discovery is a matter of empirical research, whether formally conducted by researchers or informally ascertained by observing real attacks.

When applying the technique of attack tree modeling to the threat of supply chain attacks, it is apparent that the risk assessment process is limited by the need for empirical observation of supply chain attack events and how they are causally related to known vulnerabilities. It seems unlikely, however, that the threat of a supplier could be captured by identifying a set of discrete attack events that the supplier might initiate because the set has no clear principle of construction. Suppliers have extensive discretion and, supposing the ability to act covertly, it is not feasible to rely on a discrete set of known attacks that a supplier might cause.

A second way in which supply chain attacks cannot be easily added to existing attack tree techniques is that suppliers have the potential to alter underlying system constraints. Traditional attackers work within the constraints of a system as it is deployed, finding unanticipated or vulnerable pathways to cause events that work to their advantage. Yet a supplier may work at a deeper systemic level, modifying system design specifications or introducing alien components. One critical avenue of a supply chain attack is defined by the ability to alter the system itself, its components and their functionality, including the addition or suppression of particular functions. This relatively undefined form for supply chain attacks poses a significant challenge for modeling and calls for a shift in perspective.

In the face of these difficulties applying attack tree techniques, we find that a model for supply chain attacks should be component-centered rather than event-centered. With a component-centered model it is possible to identify the suppliers that are the source of the risk under investigation, without artificially limiting the scope of their activity to some set of events that a supplier may cause. With components and suppliers as core elements, a risk model will consider the probabilities of security failures that propagate through a system. This kind of model will bear a close resemblance to classical system reliability analysis, augmented with appropriate nodes for suppliers ~\cite{rausand2003system, contini2011analysis}.

The most difficult aspect of constructing such a model for an existing system is ascertaining how component nodes are related with respect to security attributes.  This process differs from classical fault tree analysis in an important way. To construct a fault tree, it is possible to begin with detailed understanding of the component functionalities and interfaces. Although this is a complex task for a system of any moderate size, in principle the information about proper functional constraints is available. In contrast, security failures may include the emergence of unknown faults that originate from undefined behavior, software bugs, and overlooked interface gaps. These issues are difficult to foresee and information about them is largely reliant on processes of empirical research. As such it may be difficult to ascertain how the security of one component affects the security of another.

When seeking to develop and apply such a model for an existing system, the possibility of errors is a constraint within which model development must occur. In some cases the chief concern will be to develop more reliable estimations of the trustworthiness of the suppliers in a system. In other cases, though, the principal challenge in producing a reliable risk assessment will involve ascertaining the proper structure for the system in question, i.e., how the security attributes of components are causally inter-related. After outlining the necessary elements of a system model suited for supply chain risk assessment, we then investigate the effects caused by these various kinds of uncertainties.

\subsection{System Graph}
The model presented here is based on a directed graph consisting of components and suppliers. After defining essential parts of this system graph, we then describe how to calculate systemic risk given such a graph.
\subsubsection{Components}
The components in a system will be denoted as the set, $C$. The set is initialized with a component that represents the system itself. A key feature of this analysis is to consider each component as simultaneously a component and a system in itself. The set, $C$, can be expanded by conducting a recursive decomposition of each element in the set, until an arbitrary level of depth is reached. The termination condition for this recursive decomposition process can be adjusted to provide more accurate analysis at the expense of greater complexity. A general rule that defines the maximal complexity that will be useful is to terminate recursive decomposition when additional iterations will no longer yield new suppliers. In other words, a component can be considered a useful unit of analysis when it is produced entirely or for the most part by a single supplier. When this rule has been followed, the assumption is warranted that the supplier of any component in the system bears full responsibility for the component and cannot pass responsibility to one of its own suppliers. This maximal complexity still entails extensive decomposition that may yield diminishing returns as components become simpler and their suppliers possess less latitude to introduce security risk. As such, the degree of depth is considered a hyper-parameter for the construction of this model.

\subsubsection{Suppliers}
Each component in the system must be assigned a supplier from the set, $S$, where $s$ is the supplier of $c$. To be the supplier of a component will be a general term for the entity that is responsible for the manufacturing of a component. While other roles may be considered relevant to supply chain risk, such as maintenance or logistics, we consider here a generalized supplier role that may be taken to represent all aspects of supply chain risk. A rule for designating a supplier will be that a supplier must have unrestricted physical access to the component. While it may be very desirable to implement transparency and accountability schemes for suppliers, we assume here that the supplier is able to act covertly if and when it desires. 
\subsubsection{Dependencies}
Edges in this graph, $E$, will be of two sorts: i) $e \in C  \times C$ and ii) $e \in S \times C$. Edges between components will be added when the security of the destination node requires the security of the source node. Security of a component is defined by a security policy, and may refer to any combination of attributes such as \{confidentiality, availability, integrity\}. 

Edges from suppliers to components are added when the supplier is the supplier of the component in question. Here it is assumed that the supplier is \textit{de facto} a security dependency of the component, such that if the supplier were compromised or malicious, the component could no longer be considered secure.
\subsubsection{Node Logic}
Similar to attack tree models, nodes in the system graph here will each possess a logic function $\ell \in \{AND,OR\}$. Input to $\ell$ will be the set of component predecessors in the system graph, denoted as $N^-_n \cap C$, or the components that are security dependencies of the node in question. The relationship of the node to its predecessors will be dependent on $\ell$. The function $L$ assigns a component $c_i$ a particular function $\ell$.
\subsubsection{Probability Values}
Nodes in the system graph are assigned probability values, $r$, that correspond to the likelihood of a security failure at the node. These risk values for components are anterior to any consideration of predecessors or suppliers, and are intended to capture the inherent possibility of a component or supplier being compromised directly. For example, a particular router may use vulnerable software and as such possess a higher risk of security failure. However as noted above, the compromise of predecessors or the component's supplier are taken to have the equivalent effect of node security failure. In the example of the router, compromising the router's supplier would be as much of a security failure as a classic exploitation of its vulnerable software.
\subsubsection{System Graph}

With the above definitions in mind, a system graph is defined as a directed graph $ G = (V, E, L, r)$ where $V = C\cup S$. A concise example is shown in Figure \ref{fig:simple_system}. Here three component nodes are related by edges between them and each component also possesses a supplier node. 

\subsection{Risk Analysis}

Systemic risk analysis is conducted on the system graph by considering the likelihood of security failures at each node and calculating the risk to the system as a whole. Drawing on Leveson's discussion of safety in \cite{leveson}, security can be considered as an emergent property in a complex system and so provision must be made to identify indicators of this property. If the system graph is a tree, the root node is an illustrative choice for an indicator. However, the system graph may not have a root node, so it will always be necessary to specify these indicators. Here we use $I_n$ as the set of indicator nodes whose security is critical to the system, and $\ell_I$ is a logic function used to aggregate the function state of the indicator nodes. System security is then defined as $\ell_I(I_n)$.
\subsubsection{Node Failure Conditions}
There are three ways any component node $c_n$ can fail: 
\begin{itemize}
\item a direct, local security failure at the node itself, with probability $r_n$,
\item a failure of its dependencies, $\ell(N^-_n)$,
\item a failure of its supplier, $s_j$ with probability $r_j$.
\end{itemize}

\begin{figure}[t]
    \centering
    \includegraphics[width=.25\textwidth]{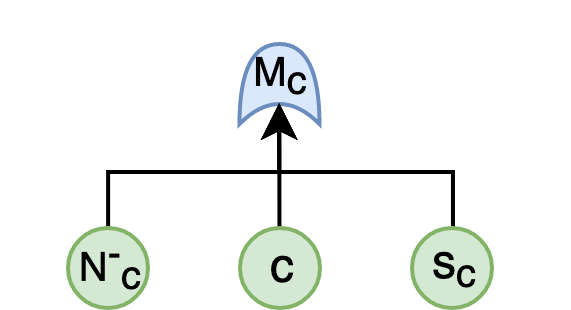}
    \caption{The conditions for node security failure involve local failures of components, failures of component dependencies, or a supplier failure.}
    \label{fig:module}
\end{figure}
These three possible failure conditions are depicted in Figure \ref{fig:module}, with the node $M_c$ representing the root of a sub-tree indicating the possible failure modes of component $c$. By substituting this module for each component node in the system graph, a complete mapping of each possible route to system security failure is possible. The root of the module will be a node with $\ell = OR$ and $r = 0$, with three predecessors: a component node with $r = r_n$, a supplier node with $r = r_j$, and a predecessor node with $\ell = L(n)$ and $r = 0$. The predecessor node will receive the incoming component edges that the original component node possessed. If the original node had no predecessors, then the predecessor node receives no incoming edges.
\subsubsection{General System Risk Calculation}
Given the above system graph, the systemic security risk may be approached with methods taken from reliability analysis. By distilling a system graph into a set of minimal cutsets, the general risk in a system is equivalent to the probability that all of the cutsets are false~\cite{rausand2003system}. Here we have relied on the MOCUS algorithm to obtain minimal cutsets on a tree with AND/OR logic \cite{fussell1974mocus}, while pointing to ongoing research in this area for algorithmic improvements \cite{lee1985fault, rauzy2003mocus, reay2002fault}. If the minimal cutsets are $W$, and $\vec{r}$ holds the probabilities of each node's failure, then the general risk is computed as: 
\begin{align}
R(\vec{r}) = 1-\prod_{w\in W}\left ( 1-\prod_{v\in w}r_v  \right).
\end{align}
\subsubsection{System Cutset Metrics}
In addition to general system risk, we note here several other metrics on a system graph that will be useful in the analysis that follows. First, we note the number of minimal cutsets, or $|W|$, which serves not as a predictor of risk but is correlated with the complexity of the system graph. While a risky system may have few cutsets that contain relatively high probability events, in such a system there are fewer variables to analyze. This factor may play a role in assessing the confidence that users may reasonably place in the utility of the results of risk analysis on such a system. 

A second, related metric is the average size of the cutsets:
\begin{align}
    \bar{W} = \frac{\sum_{w \in W} |w|}{|W|}
\end{align}
Where $\bar{W}$ is low, fewer events must occur for system failure. While, again, this cannot be predictive of risk in general without referring to the likelihood of the events in question, the metric indicates the complexity of the system.

Lastly, we define the Jaccard distance of the cutsets $W$ and $W'$ to be:
\begin{align}
    J(W, W') = 1 - \frac{|W \cap W'|} {|W \cup W'|}
\end{align}
As a measure of set similarity, the Jaccard distance will be useful in comparing two systems. Where the distance is very low, the two systems have very similar security failure conditions. Analysis and mitigation, then, of one such system will be transferable to the other with relatively high utility.

Although additions to this system model expand the model's ability to analyze risks in multi-layer networks with suppliers possessing dependencies among themselves that introduce further complexity, we present here a sufficiently developed explanation to evaluate the performance of the model under various realistic uncertainties.

\section{Uncertainties in Model Development}

Although the model we have developed bears a resemblance to established methods of system reliability analysis, the problem domains of reliability and security differ sufficiently to warrant a discussion of limits and challenges to the use of this model in practice. While certain of these challenges may be overcome through developments in methodology, others may point to limits within which the problem of supply chain security analysis must be conducted. In the analysis that follows we identify major challenges to the accurate construction of this model for a real system. After discussing these challenges in general, we illustrate the effect of four kinds of uncertainties using a case study.

\subsection{Parametric Uncertainties in Probability Estimates}

The first major area of difficulty in the use of such a model is obtaining accurate probability estimates for basic events such as component security failures and, more critically, supplier security failures. Estimating the likelihood of a supplier being compromised or being covertly malicious is a problem involving considerable difficulty. On the assumption that any compromise or malicious act will eventually be detected and attributed accurately, the accuracy of risk values will generally increase as this information is incorporated into assessed likelihoods. We consider the problem of estimating accurate risk values to be best approached through the development and use of heuristics and metrics together with information gathering and regular assessments. If accuracy is a limitation here then it is one that system design and use must accommodate.

\subsection{Structural Modeling Uncertainties}

A second source of uncertainty lies in the possibility that sources of risk are simply omitted from the system model, i.e., that some set of nodes or edges that \emph{should} be in the system graph are not included. We call these structural uncertainties, and define them as a modeling choice that has some effect on the set of minimal cutsets. Therefore the three kinds of uncertainties here will include those related to nodes, edges, and node logic functions. Being uncertain about the structure of a system could easily be a matter of neglect or oversight, but may just as well be a result of a complexity in system design that lies outside the reasonable purview of those building the model. In the case of both an inaccurate probability value and a structural modeling divergence, the calculated systemic risk value does not correspond to the real system risk. In the case studies that follow, we seek to illustrate the observation that structural uncertainties pose a significant challenge to accurate modeling and merit priority over improvements in accurate probability estimations.

\section{Uncertainty Case Studies}
In the case studies that follow, we first present a ground truth scenario that is intended to represent an ideal system graph constructed to model the system in question. Following this, we discuss four kinds of uncertainties and illustrate the possible effect of each by comparing the results of risk analysis after each error with the ground truth scenario.

\subsection{Case 0: Ground Truth}

\begin{figure*}
    \centering
    \includegraphics[width=.7\textwidth]{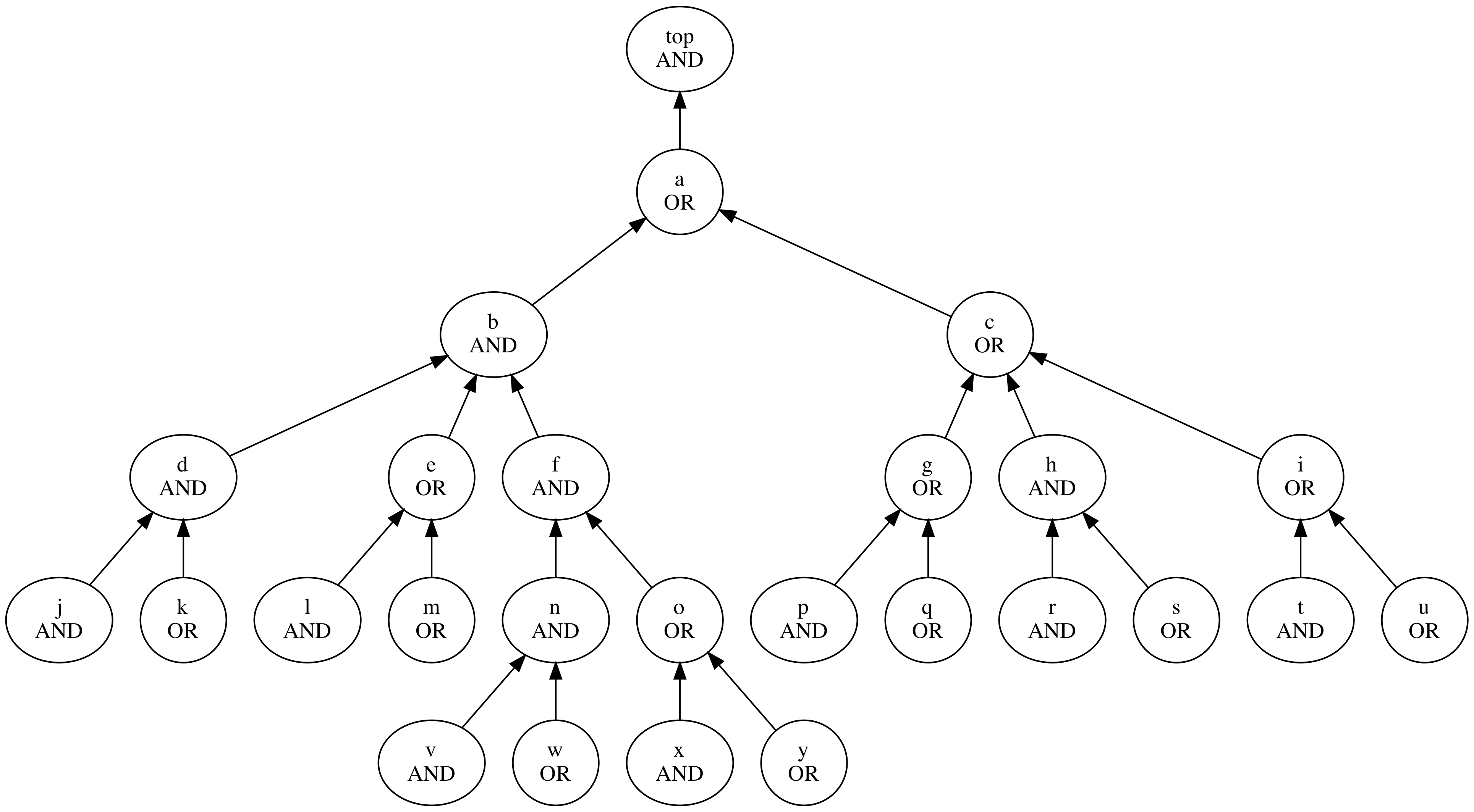}
    \caption{System Graph for Case 0: Uncertainty scenarios will be examined with reference to this as the ground truth scenario. System security is represented by the top node, and node failures that constitute minimal cutsets will cause a failure of the top node.}
    \label{fig:ground}
\end{figure*}
The system graph for this case study is shown in Figure \ref{fig:ground}. It possesses a tree structure with twenty-five nodes and with roughly equal numbers of AND and OR nodes distributed throughout the graph. We have chosen the tree structure to provide the basis of these examples because of its resemblance to classical fault trees. In practice the structure could vary widely. However, for the purpose of this study a tree structure seems likely to provide a suitable basis for generalization. 
\begin{table}
\centering
\caption{Minimal Cutsets for Case 0: Each cell contains one minimal cutset such that a failure of every node in the cutset entails a system security failure.}
\begin{tabular}{|l|l|l|}
\hline

\{a\} & \{b\} & \{c\} \\ \hline 

\{g\} & \{h\} & \{i\} \\ \hline 

\{p\} & \{q\} & \{t\} \\ \hline 

\{u\} & \{r,s\} & \{d,e,f\} \\ \hline 

\{d,f,l\} & \{d,f,m\} & \{d,e,n,o\} \\ \hline 

\{d,e,n,x\} & \{d,e,n,y\} & \{d,l,n,o\} \\ \hline 

\{d,l,n,x\} & \{d,l,n,y\} & \{d,m,n,o\} \\ \hline 

\{d,m,n,x\} & \{d,m,n,y\} & \{e,f,j,k\} \\ \hline 

\{f,j,k,l\} & \{f,j,k,m\} & \{d,e,o,v,w\} \\ \hline 

\{d,e,v,w,x\} & \{d,e,v,w,y\} & \{d,l,o,v,w\} \\ \hline 

\{d,l,v,w,x\} & \{d,l,v,w,y\} & \{d,m,o,v,w\} \\ \hline 

\{d,m,v,w,x\} & \{d,m,v,w,y\} & \{e,j,k,n,o\} \\ \hline 

\{e,j,k,n,x\} & \{e,j,k,n,y\} & \{j,k,l,n,o\} \\ \hline 

\{j,k,l,n,x\} & \{j,k,l,n,y\} & \{j,k,m,n,o\} \\ \hline 

\{j,k,m,n,x\} & \{j,k,m,n,y\} & \{e,j,k,o,v,w\} \\ \hline 

\{e,j,k,v,w,x\} & \{e,j,k,v,w,y\} & \{j,k,l,o,v,w\} \\ \hline 

\{j,k,l,v,w,x\} & \{j,k,l,v,w,y\} & \{j,k,m,o,v,w\} \\ \hline 

\{j,k,m,v,w,x\} & \{j,k,m,v,w,y\} & \\ \hline
\end{tabular}
\label{ground_cutsets}
\end{table}

\begin{table}
\centering
\caption{Results for Case 0, Ground Truth}
\begin{tabular}{|l|l|}
\hline

$|W|$ & 53 \\ \hline
$avg(|w|)\, \forall w \in W$ & 4.018868 \\ \hline
$J(W,W')$ & 0.0 \\ \hline
$Risk$ & 0.403032 \\ \hline
$\Delta Risk$ & 0 \\ \hline

\end{tabular}
\label{ground_results}
\end{table}

The minimal cutsets of this system are shown in Table \ref{ground_cutsets}. Each cell contains a set of nodes identified by alphabet letter, where the security failure of all nodes in the set represents a security failure of the top node. To compute a risk value for the system, we provide sample component risk values such that each component has a risk of failure of 0.05. Essential metrics for this system are found in Table \ref{ground_results}.

\begin{figure}[h!]
    \centering
    \includegraphics[width=.45\textwidth]{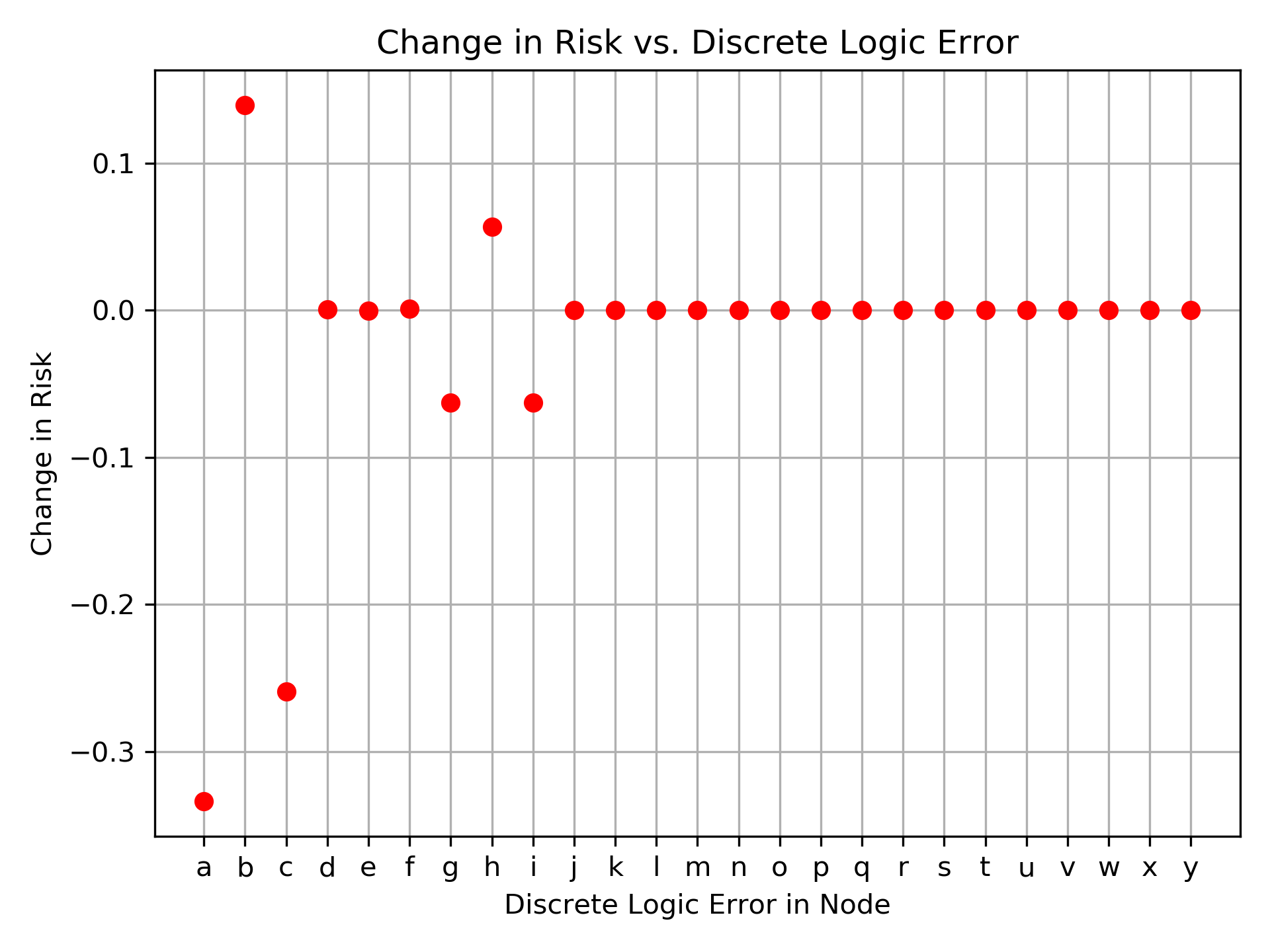}
    \caption{Case 1 Results: When node logic is subject to discrete error, systemic risk values vary widely, but with a magnitude related to node height.}
    \label{fig:case1_all}
\end{figure}

\begin{table}[h!]
\centering
\caption{Results for Case 1, Logic Uncertainty in $c$}
\begin{tabular}{|l|l|}
\hline

$|W|$ & 63 \\ \hline
$avg(|w|)\, \forall w \in W$ & 4.238095 \\ \hline
$J(W,W')$ & 0.366197\\ \hline
$Risk$ & 0.144027 \\ \hline
$\Delta Risk$ & -0.259005 \\ \hline

\end{tabular}
\label{case1_results_c}
\end{table}

\begin{figure}[t]
    \centering
    \includegraphics[width=.45\textwidth]{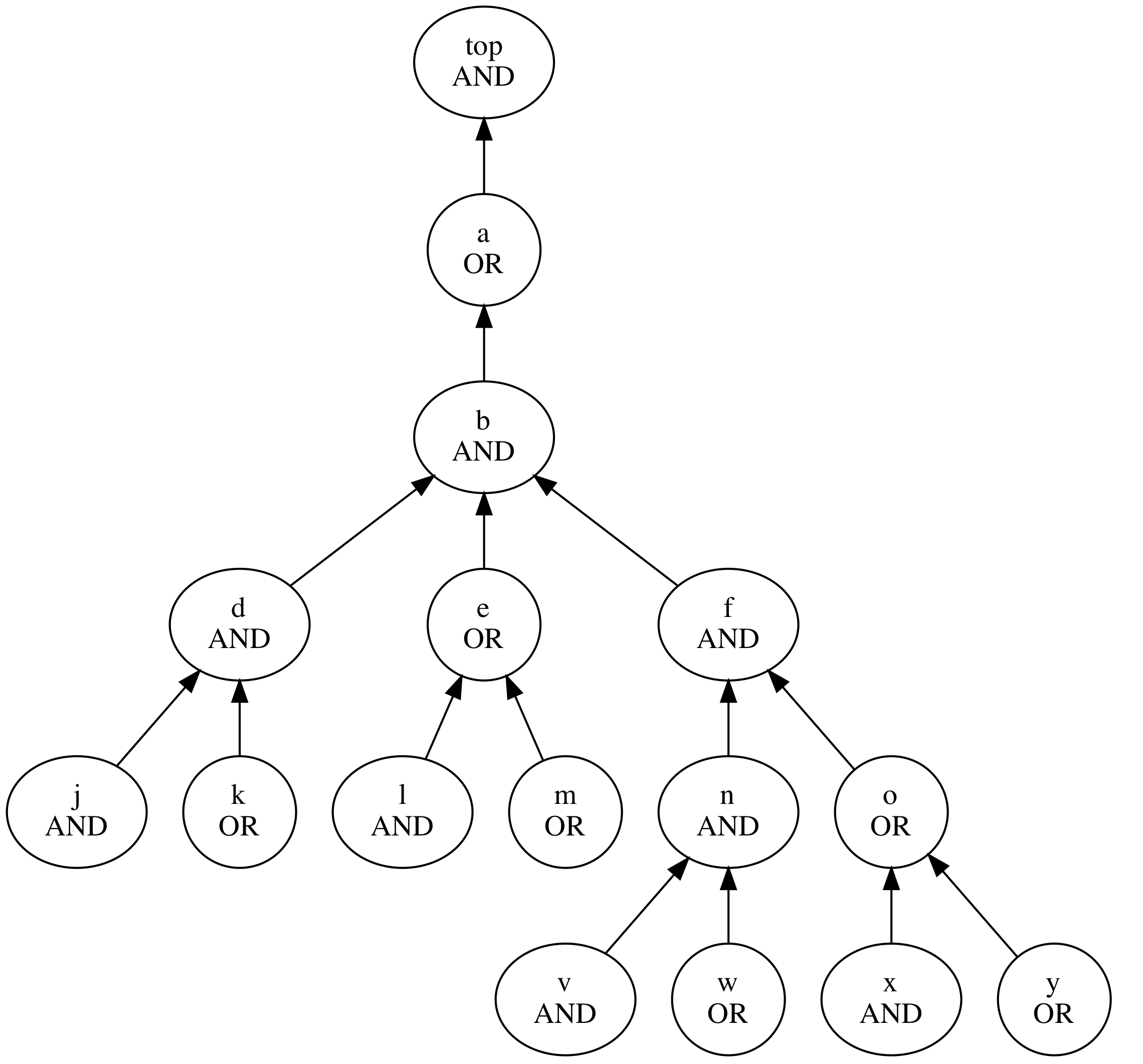}
    \caption{System Graph for Case 2, illustrating an erroneous omission of the component at node $c$ as well as its sub-components.}
    \label{fig:case2_graph_c}
\end{figure}
\vspace{-0.0in}

\subsection{Case 1: Uncertainty of Single Node Logic}

In this first uncertainty scenario, the logic type of a single node will be modified to represent the misclassification of a node with regard to its predecessors. A full treatment of the case of a single logic error suggests investigating the effect of this error on any given node in the graph. While this would indeed yield a more thorough understanding, such a generalized study would be of limited value without operating on a generalized graph. In lieu of this theoretical exercise, here we present the result of analysis when various nodes in the case study are mistaken. We chose nodes \emph{c} and \emph{b}, where the analysis will be conducted on the graph after each node's logic function $\ell_n \in \{AND,OR\}$ has been substituted for the opposite type. Descriptively, this entails an error in recognizing the way that components \emph{g, h, i} and their predecessors affect the security of component \emph{c}, with an analogous error for node \emph{b}. While the ground truth scenario includes a more risk-amplifying relationship, where any of \emph{g, h, i} can cause \emph{c} to fail, the situation studied in Case 1 is that the model designer considers \emph{c} to fail when all of \emph{g, h, i} have failed. As such, we expect that this analysis will result in an erroneously low risk assessment when node \emph{c} has been modified, while the opposite will be the case for node \emph{b}.

Detailed results for Case 1 on node \emph{c} are shown in Table \ref{case1_results_c}, including a modest rise in the number of cutsets as well as their average size. The Jaccard distance from the ground truth is 0.366, indicating a probability of roughly 1/3 that a cutset in either case is not shared between the two. Risk, as expected, has dropped by 0.259. To contrast, we present the results for changing node \emph{b} in Table \ref{case1_results_b}. This single node logic error results in a significant Jaccard distance of 0.83, while raising systemic risk by 0.139. Finally, Figure \ref{fig:case1_all} shows the effect of a logic error at each node in the system. Because many nodes are leaf nodes possessing no dependencies, the nature of the logic function at the node is irrelevant to systemic risk. Similarly, we note a general correlation between the magnitude of the change in risk and the height of the node in question.

\vspace{-0.0in}

\subsection{Case 2: Uncertainty of Node Omission}

When constructing a model, it may easily occur that a component is overlooked and omitted from the model. Especially as complex systems involve many layers of components, there will be some uncertainty concerning whether important nodes have been omitted. To capture this uncertainty, we test here the result of deleting a node from the system graph. As in Case 1, much of the effect of such an error will depend on the topography of the system graph as well as the location of the node omitted. When omitting a node, we consider it necessary to omit also the children of the node that become disconnected from the graph as a result. This choice reflects the likelihood that in overlooking a component, its subcomponents or dependencies will also be overlooked. We test here the omission of a mid-level node \emph{f} and, separately, a higher-level node \emph{c}. Figure \ref{fig:case2_graph_c} shows the modified system graph with node \emph{c} omitted.
\vspace{-0.0in}

\begin{figure}
    \centering
    \includegraphics[width=.45\textwidth]{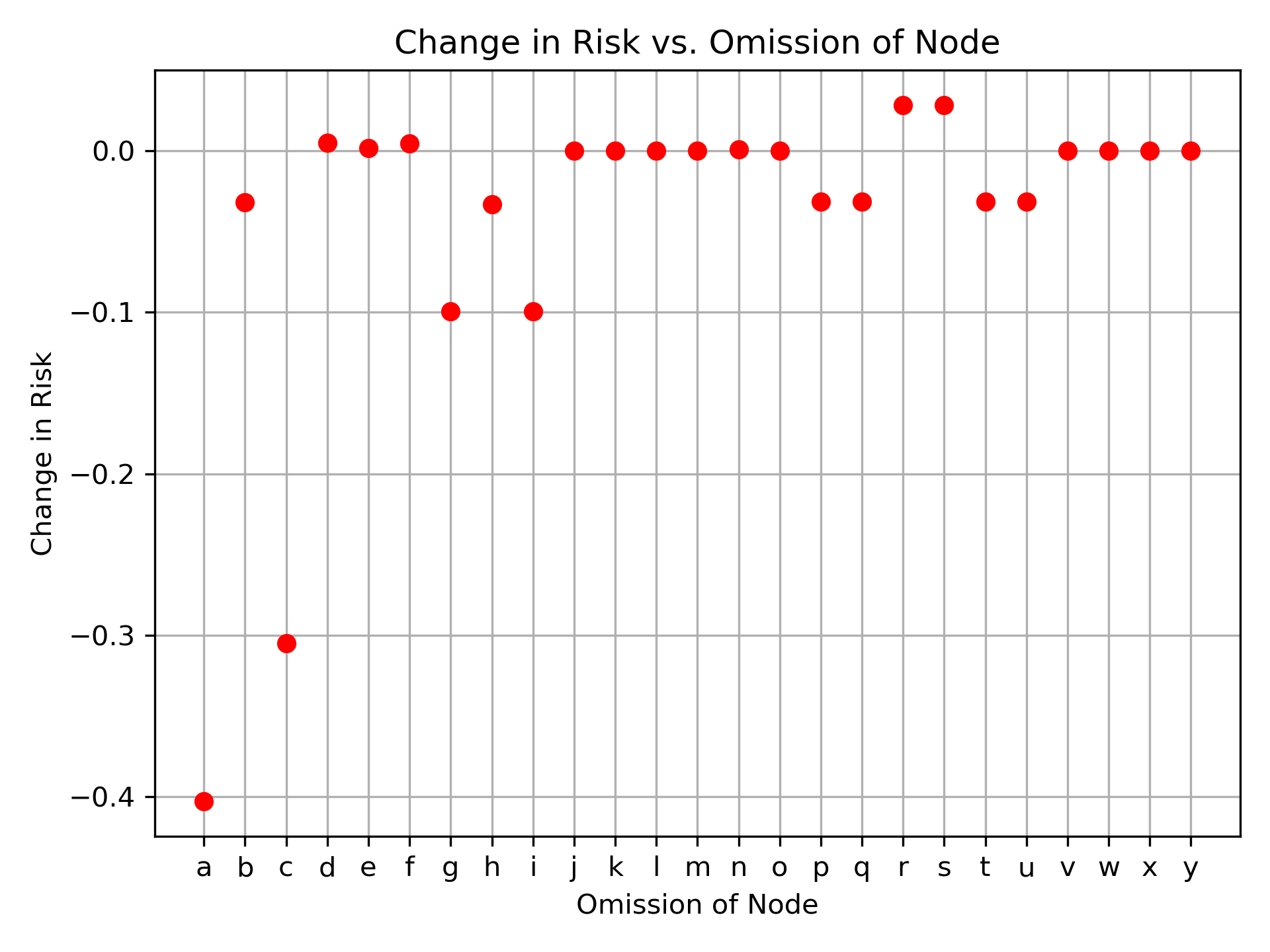}
    \caption{Case 2 Results, showing the change in system risk when a single node and the sub-tree rooted at the node is omitted. The magnitude of change in risk is not strictly correlated with node height.}
    \label{fig:case2_all}
\end{figure}

Detailed results are shown in Tables \ref{case2_results_f} for the omission of node \emph{f} and \ref{case2_results_c} for the omission of node \emph{c}, while a survey of the resulting change in risk for each node's omission is shown in Figure \ref{fig:case2_all}. It is pertinent to note the lack of correlation between the Jaccard distance of the minimal cutsets and the change in systemic risk. While omitting node \emph{f} yields a very significant distance between the cutsets (0.81), the change in risk is minimal (0.004). By contrast, omitting node \emph{c} results in a smaller Jaccard distance (0.17) but a very large decrease in risk (0.305). This volatility in modeling results points toward the importance of component level analysis in understanding supply chain risk. We also note the lack of correlation here between the change in risk and the height of the node in question. Omitting node \emph{c} has a rather large effect, whereas node \emph{b}, with the same height, has a very small effect.

\begin{table}[h!]
\centering
\caption{Results for Case 1, Logic Uncertainty in $b$}
\begin{tabular}{|l|l|}
\hline

$|W|$ & 23 \\ \hline
$avg(|w|)\, \forall w \in W$ & 1.478261 \\ \hline
$J(W,W')$ & 0.830769 \\ \hline
$Risk$ & 0.542643 \\ \hline
$\Delta Risk$ & 0.139611 \\ \hline

\end{tabular}
\label{case1_results_b}
\end{table}

\begin{table}[h!]
\centering
\caption{Results for Case 2, Node Omission in $f$}
\begin{tabular}{|l|l|}
\hline

$|W|$ & 17 \\ \hline
$avg(|w|)\, \forall w \in W$ & 1.588235 \\ \hline
$J(W,W')$ & 0.813559 \\ \hline
$Risk$ & 0.407450 \\ \hline
$\Delta Risk$ & 0.004418 \\ \hline

\end{tabular}
\label{case2_results_f}
\end{table}

\begin{table}[h!]
\centering
\caption{Results for Case 2, Node Omission in $c$}
\begin{tabular}{|l|l|}
\hline

$|W|$ & 44 \\ \hline
$avg(|w|)\, \forall w \in W$ & 4.613636 \\ \hline
$J(W,W')$ & 0.169811 \\ \hline
$Risk$ & 0.097911 \\ \hline
$\Delta Risk$ & 
-0.305121 \\ \hline

\end{tabular}
\label{case2_results_c}
\end{table}

\subsection{Case 3: Uncertainty in Edge Placement}

The scenario captured as uncertainty in the placement of a single edge will be when a component node is successfully identified but it is mistaken how the node is related to other nodes in the system. As such, an edge error entails no change in the number of connected nodes in the graph. There may be a large number of possibilities that are plausible ways an edge might be mistakenly placed. As such it will be difficult to examine all the nodes as was done in the previous two cases. We present detailed results of two different edge errors. First, we remove edge $\langle d,b \rangle$ and substitute it for the edge $\langle d, e\rangle$. The results of this error are shown in Table \ref{case3_results_db}. To contrast, we also investigate the change of edge $\langle h,c \rangle$ to $\langle h,g \rangle$, the results of which are shown in Table \ref{case3_results_hc}

In the first examined edge modification we find a minimal change in risk despite a large distance between the cutsets. In contrast, a change in the edge $\langle h,c \rangle$ yields the identical minimal cutsets, owing to the nature of the original parent node's logic. 

\begin{table}[h]
\centering
\caption{Results for Case 3, Edge $\langle d,b \rangle \to \langle d, e\rangle$}
\begin{tabular}{|l|l|}
\hline

$|W|$ & 46 \\ \hline
$avg(|w|)\, \forall w \in W$ & 2.913043 \\ \hline
$J(W,W')$ & 0.875 \\ \hline
$Risk$ & 0.409726 \\ \hline
$\Delta Risk$ & 
0.006694 \\ \hline

\end{tabular}
\label{case3_results_db}
\end{table}

\begin{table}[h]
\centering
\caption{Results for Case 3, Edge $\langle h,c \rangle \to \langle h, g\rangle$}
\begin{tabular}{|l|l|}
\hline

$|W|$ & 53 \\ \hline
$avg(|w|)\, \forall w \in W$ & 4.018868 \\ \hline
$J(W,W')$ & 0.0 \\ \hline
$Risk$ & 0.403032 \\ \hline
$\Delta Risk$ & 
0.0 \\ \hline

\end{tabular}
\label{case3_results_hc}
\end{table}

\begin{table}[h]
\centering
\caption{Results for Case 4}
\begin{tabular}{|l|l|l|l|l|}
\hline

$e$ & 0.02 & 0.05 & 0.10 & 0.50 \\ \hline
$Risk$ & 0.409364 & 0.418751 & 0.434108 & 0.544767 \\ \hline
$\Delta Risk$ & 0.006332 & 0.015719 & 0.031076 & 0.141735\\ \hline

\end{tabular}
\label{case4_results}
\end{table}

\subsection{Case 4: Uncertainty in Probability Values}

\begin{figure}
    \centering
    \includegraphics[width=.45\textwidth]{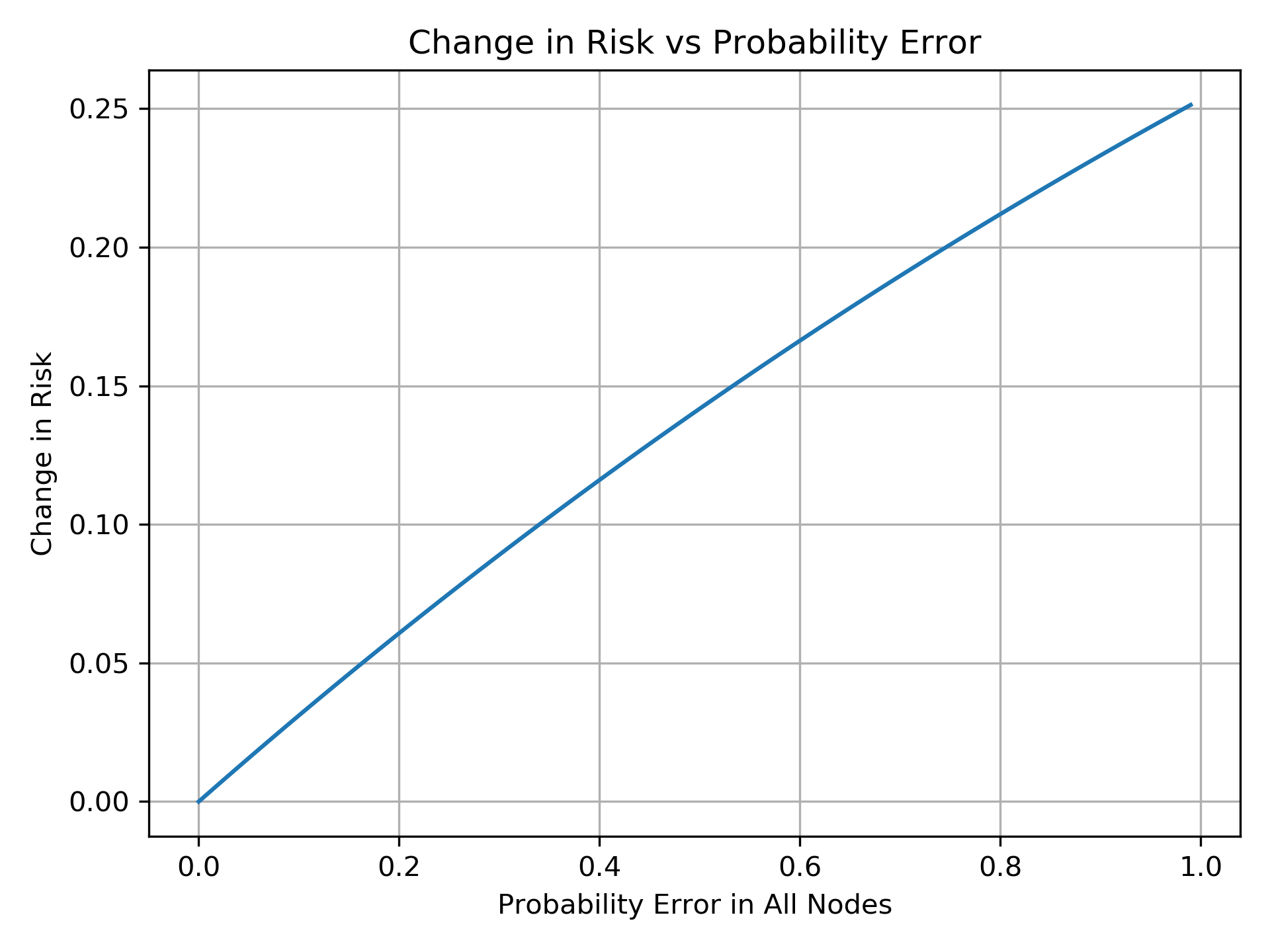}
    \caption{Case 4 Results, where the probabilities of all events are adjusted by increasing margins of error. Elevated probability values entail a linear increase in risk, but only high margins of error are comparable to many structural uncertainties.}
    \label{fig:case4_all}
\end{figure}

After having explored the various kinds of structural uncertainties and illustrated their potential effects in particular cases, we examine here the contrasting effects of uncertainties in the estimation of probability values. These probability values are critical points of data without which a model cannot approximate the real risk in a system. Yet because of the difficulty of obtaining these values with accuracy and confidence, we examine the general effect of various margins of error. When applying each margin of error, $0<e\leq 1$, the adjustment is made by adding $er_i$ to element $i$ of vector $\vec{r}$. With this adjusted vector, the general risk function is calculated. Because this class of errors involves no change to the number or identity of cutsets, we only compare the resulting risk value to the ground truth scenario presented above. In Table \ref{case4_results}, we show the effect on risk analysis of four margins of error, $e$: 2\%, 5\%, 10\% and 50\%. Figure \ref{fig:case4_all} shows a range of errors and the resulting change in systemic risk. 

We note that errors are calculated with reference to the ground truth scenario, where the probability of each event is 0.05. As such, the maximum error shown in Figure \ref{fig:case4_all}, 100\%, results in an adjusted probability of 0.10. Likewise, we apply this margin of error to every node in system graph. While more complex or drastic scenarios can be imagined, these high error rates are sufficient to illustrate the relative impact of uncertainties of different kinds.

\section{Conclusion and Future Work}
In this paper we have presented a modification of attack tree modeling suited for the analysis of supply chain risks with the primary intention of investigating the practical utility of such a model when faced with inevitable difficulties in obtaining accurate data describing complex ICT and IoT critical infrastructure systems. 

The preceding case studies have depicted various possible error scenarios that may be encountered while applying this modeling technique to an existing system. Using a particular system graph, we have illustrated these error scenarios with the help of several examples. Although caution is warranted when approaching a problem of this complexity from particular case studies, we use the results shown here to highlight the importance of structural errors in comparison to errors in obtaining accurate probability estimates. If the security of components and the trustworthiness of suppliers can be estimated to within 50\% accuracy, our results show a maximum possible error in risk assessment of 14\%. This is a significant change in risk, but equal or far greater discrepancies are found with a wide variety of discrete structural errors. Mistaking a single node's logic function, if it is a systemically important node, may produce double the change in assessed risk. Similar discrepancies are found with single node omissions or mistaken edges. Generally, the nodes with greater height in the system graph are more conducive to yielding larger discrepancies in systemic risk. 

From these observations certain principles in the practical development of supply chain risk assessments may be suggested. The following preferences summarize the conclusions of this study.
\begin{itemize}
    \item \textbf{Structure over magnitude:} Given the scarcity of resources available to conduct risk assessments, and the possible impact of errors of various kinds, we suggest significant attention be given to accurate structural modeling. While efforts to obtain accurate magnitudes in risk and trust values are certainly important, the development of accurate structural models for the ways in which components relate to each other as security dependencies should usually be prioritized.
    \item \textbf{Height over depth:} At higher levels of systemic analysis, accuracy in structural modeling should take unambiguous priority. Structural errors in the critical window of 2-3 hops from the top event have the potential to make extraordinary differences in modeling results. Extensive and accurate modeling of depth into a system may be helpful, but it is less important than ensuring accuracy in this critical window. The difficulty of obtaining accurate modeling at lower levels is matched by a decrease in the impact of possible errors. As such, less effort should be expended on components at these lower levels of a system. 
\end{itemize}

\small
\vspace{-0.0in}
\bibliographystyle{IEEEtran}
\bibliography{references}

\end{document}